\documentclass[conference]{IEEEtran}
\usepackage{subcaption}
\usepackage{caption}
\usepackage{listings}
\usepackage{tikz}
\usepackage{amsmath}
\usepackage{multirow}
\usepackage{tabulary}
\usepackage{tikz,pgfplots,pgfplotstable,tikzscale,siunitx, xcolor}
\usetikzlibrary{patterns}
\usepackage{filecontents}
\usepackage{graphicx}
\usepackage[algo2e,ruled,vlined]{algorithm2e}
\usepackage{enumitem}
\usepackage{stfloats}
\usetikzlibrary{pgfplots.groupplots}
\usepackage{cite}
\usepackage{amsmath,amssymb,amsfonts}
\usepackage{algorithmic}
\usepackage{graphicx}
\usepackage{textcomp}
\usepackage{xcolor}
\usepackage[breaklinks=true]{hyperref}
\def\BibTeX{{\rm B\kern-.05em{\sc i\kern-.025em b}\kern-.08em
    T\kern-.1667em\lower.7ex\hbox{E}\kern-.125emX}}
\begin{document}

\title{FPGA-based Hyrbid Memory Emulation System
}
\author{\IEEEauthorblockN{Fei Wen, Mian Qin, Paul V. Gratz, A.L.Narasimha Reddy}
\IEEEauthorblockA{
Department of Electrical and Computer Engineering\\
Texas A\&M University\\
College Station, TX 77843\\
 {\tt \{fwen, celery1124, pgratz, reddy\}@tamu.edu}}}
\maketitle
\thispagestyle{plain}
\pagestyle{plain}
\begin{abstract}
  Hybrid memory systems, comprised of emerging non-volatile memory
  (NVM) and DRAM, have been proposed to address the growing memory
  demand of applications. Emerging NVM technologies, such as
  phase-change memories (PCM), memristor, and 3D XPoint, have higher
  capacity density, minimal static power consumption and lower cost
  per GB. However, NVM has longer access latency and limited write
  endurance as opposed to DRAM. The different characteristics of two
  memory classes point towards the design of hybrid memory systems
  containing multiple classes of main memory.
  
  In the iterative and incremental development of new architectures,
  the timeliness of simulation completion is critical to project
  progression.  Hence, a highly efficient simulation method is needed
  to evaluate the performance of different hybrid memory system
  designs.  Design exploration for hybrid memory systems is challenging, because it requires emulation of the
  full system stack, including the OS, memory controller, and
  interconnect. Moreover, benchmark applications for memory performance test typically have much larger working sets, thus taking even longer simulation warm-up period.
  
  In this paper, we propose a FPGA-based hybrid memory system
  emulation platform.  We target at the mobile computing system, which
  is sensitive to energy consumption and is likely to adopt NVM for
  its power efficiency.  Here, because the focus of our platform is on
  the design of the hybrid memory system, we leverage the on-board
  hard IP ARM processors to both improve simulation performance while
  improving accuracy of the results.  Thus, users can implement their
  data placement/migration policies with the FPGA logic elements and
  evaluate new designs quickly and effectively.  Results show that our
  emulation platform provides a speedup of 9280x in simulation time
  compared to the software counterpart Gem5.
\end{abstract}
\section{Introduction}
\label{sec:intro}
In the era of big data, the memory footprint of applications has
expanded enormously. DRAM energy consumption grows linearly with its
capacity, as DRAM cells constantly draw energy to refresh its stored
data.  The non-volatile-memory (NVM) technologies, such as Intel 3D
Xpoint~\cite{intel-3dxpoint}, memristor~\cite{memristor}, and
Phase-change-memory (PCM)~\cite{PCM} have emerged as alternative
memory technologies to DRAM.  These new memory technologies promise an
order of magnitude higher density~\cite{techinsights} and minimal
background power consumption.  Their access delay is typically within
one order of magnitude larger than that of DRAM while the write access
has significantly higher overheads.  
\begin{table*}[hbt]
  \centering
  \small
\caption{Approximate Performance Comparison of Different Memory
  Technologies\cite{NVM1,NVM2,yang:2012} }
\begin{tabular}{c|c|c|c|c|c|c}
\hline
Technology& HDD & FLASH & 3D XPoint & DRAM & STT-RAM & MRAM\\
\hline
Read Latency & 5ms & $100\mu s$  & 50 - 150ns & 50ns & 20ns & 20ns\\
Write Latency & 5ms & $100\mu s$  & 50 - 500ns & 50ns & 20ns & 20ns\\
Endurance (Cycles) & $>10^{15}$ & $10^{4}$ & $10^9$ & $>10^{16}$ & $>10^{16}$ & $>10^{15}$\\
\$ per GB & 0.025-0.5 & 0.25-0.83 & 6.5~\cite{nvm_price} & 5.3-8 & N/A & N/A\\
Cell Size & N/A & $4-6F^2$ & $4.5F^2$ ~\cite{techinsights} & $10F^2$ & $6-20F^2$ & $25F^2$\\
\hline 
\end{tabular}
\label{tab:nvms}
\end{table*}
Such different characteristics
require a significant redesign of the memory system architecture, in
both data management policies and mechanisms.  Hybrid memory systems,
comprised of emerging non-volatile memory (NVM) and DRAM, in
particular, look promising for future system design.  As to date these
new NVM technologies are only beginning to become widely available,
many open questions have yet to be worked out in management policy and
implementation.  Evaluating many alternative approaches in hybrid
memory design is essential for future system design.

Evaluating hybrid memory systems presents several unique challenges
because we aim to test the whole system stack, comprising not only the
CPU, but also the memory controller, memory devices and the
interconnections.  Further, since the study is focused on main memory,
accurate modeling of DRAM is required.  Much of the prior work in the
processor memory domain relies upon software simulation as the primary
evaluation framework with tools such as Champsim~\cite{champsim} and
gem5~\cite{gem5}. However, cycle-level software simulators that meets our requirement of system complexity and flexibility, impose huge simulation
time slow-downs versus real hardware. Moreover, there are often
questions of the degree of fidelity of the outcome of arbitrary
additions to software simulators~\cite{7155440}.

In this paper, we propose an evaluation system with the flexibility to
allow the user to define their own hybrid memory management policy and
with the performance to run full applications at near native speed.
In particular our evaluation platform leverages an FPGA platform with
a hard IP ARM Cortex A57 processors and two memory controllers
connecting the FPGA to DRAM DIMMS.  This platform with our framework
allows the hybrid memory management unit to be implemented in the FPGA
in RTL, where the real application memory requests on the host ARM
cores are redirected to memory DIMMs connected to the FPGA. Thus we
successfully decouple the design under test, i.e, the memory
management policies, from the detailed modelling of memory device. Our
FPGA-based emulation platform provides the flexibility to develop and
test sophisticated memory management policies, while its hardware-like
nature provides near-native simulation speed.

In this paper we detail the evaluation platform and compare its
performance to comparable simulation based approaches to hybrid memory
management simulation.  In particular, experimental results show
that our platform provides 2286x speedup compared to Champsim, and 9280x
speedup over Gem5.
\section{Background and Motivation}
Hybrid memories, comprised of DRAM and NVM, have recently become a hot
research topic, with numerous works~\cite{Hassan:2015,span,
  Liu:2017,ramos, CSu} in this domain.  For design evaluation, these
works predominantly use either software-based platform simulation,
with simulator runtime limiting the workloads that can be examined, or
they use analytical modeling, which has a large impact on accuracy.

In other domains, FPGA-accelerated simulators have demonstrated high
time efficiency and accuracy, far beyond the scope of typical
software-based alternatives.  Below we summarize the main advantages
of FPGA-accelerated simulators and show how this approach can be
applied to hybrid memory emulation system.


\subsection{Speed and Accuracy} 
FPGAs are composed of millions of lookup tables, each of them
programmable to fulfill certain logic functions.  Owing to its
hardware nature, FPGAs are great for parallel tasks and concurrent
execution.  Software-simulators, which are by nature sequentially
executed, have hit the simulation wall; a phenomenon that the
simulation efficiency declines as the target hardware system becomes
more complex. While software-based simulators are easier to build,
they are slower, and less accurate. Therefore several prior works have
turned to FPGA-accelerated simulators.

Chung \emph{et al.}~\cite{Chung:PROTOFLEX} introduced Protoflex for
multiprocessor simulation, which greatly reduced the simulation
duration for complex commercial applications.
ReSim~\cite{Fytraki:ReSim} is a trace-driven hardware simulator using
FPGA. It improves the timing accuracy of simulation results, while
lowering both hardware and time cost.\par
When it comes to studies on memory systems, the efficiency problem deteriorates for software-based simulation. First, the
traffic events in interconnections are highly unpredictable, thus the
latency shows a wide variability. In contrast, the CPU events can be
modeled by a fixed number of pipeline stages. Thus, prior
works~\cite{Aport} simply use a one-cycle ``magic" memory to avoid
modeling a realistic memory hierarchy, due to the timing complexity.
\par
Memory performance benchmark applications leverage large working sets to exert pressure on memory systems. Multi-GB data of the working set must be warmed up before we could exclude the transient performance fluctuation at cold start. Simulation acceleration techniques such as SimPoint and sampling, also require repetitive memory warm-up each time before we can obtain correct and representative sample. All these factors makes the simulation time prohibitive long with traditional software simulators. 
Hence, an efficient FPGA-based emulation platform is highly desired
for studies on hybrid memory systems.

\subsection{Flexibility and Completeness}
While FPGA-based simulators are primarily recognized for their high
simulation efficiency and accuracy, they also provide unique scopes to
other system metrics.  PrEsto~\cite{Sunwoo:PrEsto} is a simulator
created by Sunwoo \emph{et al.} for power consumption estimation. It's
able to predict cycle-by-cycle power dissipation, at a speed several
orders of magnitude faster than the software-based competitors. Such
detailed dynamic power information is essential for modern processor
designs.  ATLAS~\cite{Njoroge:ATLAS} is the first FPGA prototyping of
a chip-multiprocessors, with hardware-level support for transactional
memories.

The programmability of FPGA is also crucial to hybrid memory system
emulation. Researchers are free to implement their own memory
management policies in the logic blocks. Further, they don't have to
deal with the modelling of memory device itself, as all memory
requests are automatically directed to the real memory DIMMs in our
platform.

The flexibility of FPGA also allow us to emulate various NMV
technologies which has different access characteristic, as shown in
Table~\ref{tab:nvms}. Our platform provides a simple way for
customizing critical parameters such as latency.  FPGA also meets the
completeness requirement for memory system evaluations: users can
easily add a variety of performance counters of their choice. For
example, we implemented counters for read/write transactions to each
memory device respectively, and obtained a fairly accurate estimate of
the dynamic power consumption.
\section{Design}
\label{sec:design}
\begin{figure}[h]
\centering
\hspace*{-0.2in}
\subfloat[Target System Architecture]{\includegraphics[width=\columnwidth]{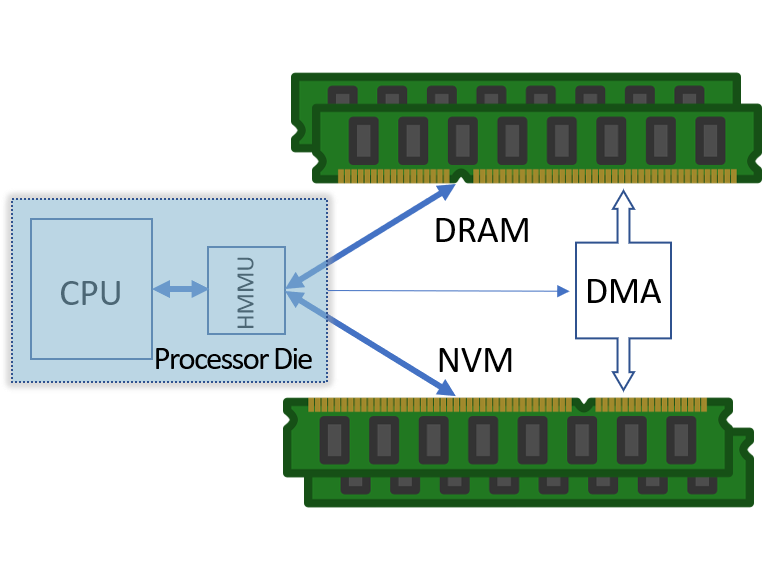}
\label{fig:target}}
\hfil
\hspace*{-0.2in}
\subfloat[Emulation System Architecture]{\includegraphics[width=\columnwidth]{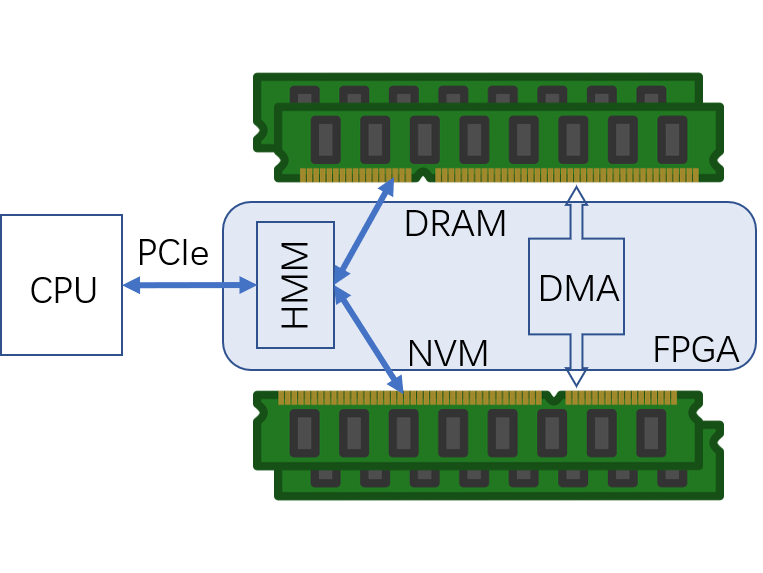}%
\label{fig:emulation}}
\caption{Target System and Emulation System Architecture}
\label{fig:architecture}
\end{figure}
Figure~\ref{fig:target} shows the target system and emulation system of our proposed scheme. The target design is the Hybrid Meomory Manger Unit (HMMU) that will be integrated to the processor die. The HMMU directly manages two memory device and receives the memory requests from the host CPU after cache filtering. 
The received memory requests are then processed based on the user-defined data placement policies, and forwarded correspondingly to either DRAM or NVM.  The HMMU also manages the data migration between DRAM and NVM, by controlling the high-bandwidth DMA engine that connects to the both memory devices.\par
In the emulation system, as illustrated in Figure~\ref{fig:emulation}, the HMMU and DMA engine are implemented in FPGA, and connects to the host via PCIe links.

\subsection{Memory Request Processing Workflow}
\begin{figure*}[h]
\centerline{\includegraphics[width=0.8\textwidth]{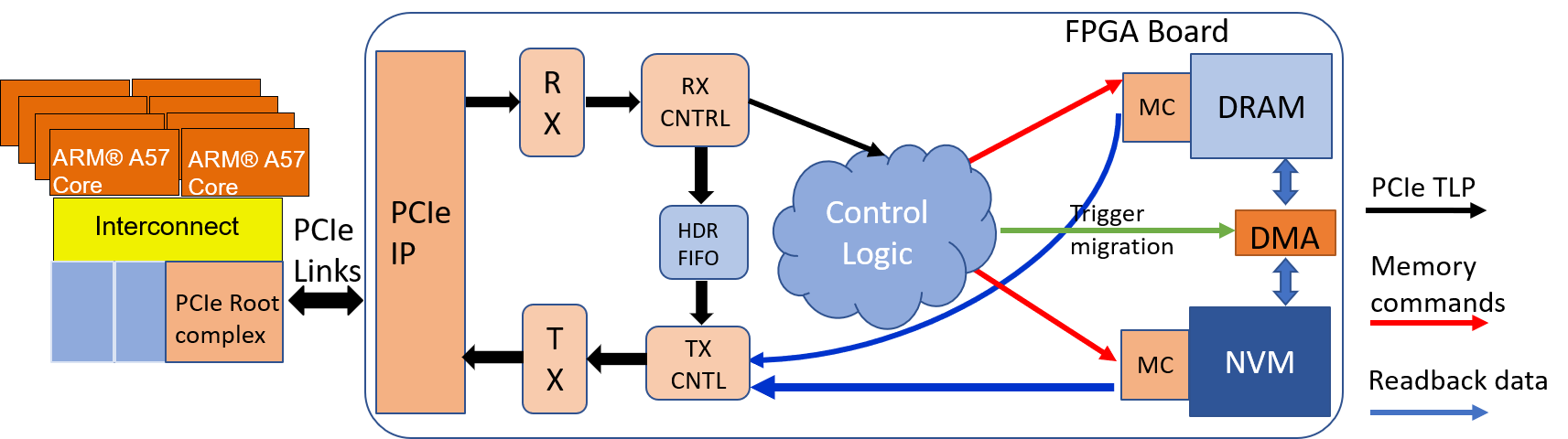}}
\caption{Request Processing Workflow}
\label{fig:flow}
\end{figure*}
Figure~\ref{fig:flow} briefly walks through the memory request processing workflow:
PCIe hard IP block receives TLPs carrying the memory requests from the host CPU, and then forwards them to the RX module of the FPGA logic.
RX Control module extracts the TLP header into the FIFO by the order they were received. Meanwhile the TLPs are also forwarded to the control logic which is highly pipelined. In the first stage, TLPs were decoded and the interpreted memory requests are populated into the following stages. Here, you can design your own memory management policies, which usually have three aspects: the memory access pattern recognition, data placement policy, and data migration policy.
We'd elaborate these policies in the later sections. The outcome of the control logic can also contain actions corresponding to these policies: 
\begin{itemize}
    \item Memory requests are forwarded to the memory controller (MC) of target device.
    \item Trigger data migration in the DMA.
\end{itemize}
The journey ends for write memory request, when they arrive at the MC and the payload data were written into target memory device.
As for the read requests, the read back data are retrieved from memory, and returned in the same order as the requests were received.
The TX CNTL module receives the read back data from both NVM and DRAM, and assembles them with the corresponding header into a complete response message. The response messages are then encapsulated into PCIe transactional layer packets (TLP) at TX and then are sent back to the CPU host by the PCIe hard IP block.
\subsection{Heterogeneity Transparency}
Although hybrid memory is composed of two separate memory devices, designer might want to hide such heterogeneity from OS, and present one single flat memory space. Thus, all the data placement and migration between DRAM and NVM becomes transparent from the user applications' perspective. This mechanism is beneficial in several aspects:
\begin{enumerate}
    \item All the data movement are executed by purely hardware-based HMMU, without interrupting the OS and user applications.
    \item Programmers don't have to learn about data allocation among separate memory devices.
    \item Compatible with legacy applications.
\end{enumerate}
There are many implementation methods to realize this design. A straightforward approach is to build another layer of address redirection table within the HMMU, where the physical address is translated to the actual memory device address. The mapping rule becomes part of the data placement policy.
\subsection{Memory Consistency}
\begin{figure}[h]
\centerline{\includegraphics[width=\columnwidth]{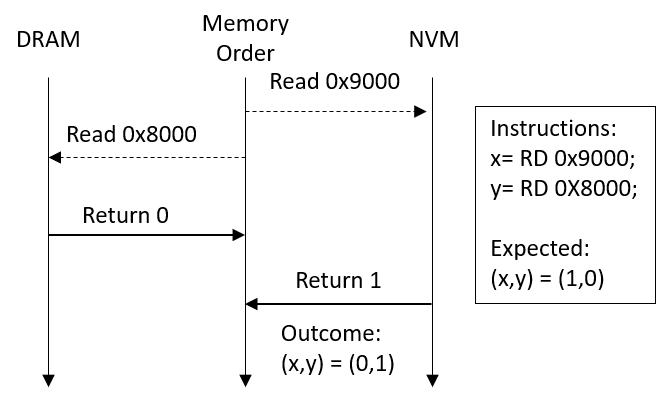}}
\caption{Memory Consistency Risk}
\label{fig:consistency}
\end{figure}

A risk of memory consistency error arises here as the memory requests are split into the two channels for NVM and DRAM, respectively. From the perspective of OS, there is only one single memory space and the OS cannot tell which memory device is the request for. Hence, a consistency error is possible as shown in Fig. ~\ref{fig:consistency}, when the data was not returned in the same order as it was requested. Such errors are prone to happen when the later request is aimed at DRAM, which has a lower access latency. We adopt a tag-matching mechanism to guarantee the consistency, while still allowing out-of-order memory media access for higher performance. In particular, we use the header information ,stored at \texttt{HDR FIFO} in Fig.~\ref{fig:flow}, as the tag to save the order of memory requests.
\subsection{DMA Engine}
To efficiently migrate data between DRAM and NVM, without interfering processor memory requests, we need to implement a dedicated DMA engine.\par
The bit width of DMA shall be compliant with the memory controller for maximum throughput. Due to the unbalanced data rates of the two memory devices, and the discrepancy of clock frequencies between the DMA module and memory controllers, an internal buffer is needed to temporarily hold the data during migration. Besides the choice of these two primary design parameters, bit width and buffer size, a major concern is how to guarantee data coherence when a memory access conflicts with ongoing DMA data migration.\par
When DMA swaps two pages, the data is transferred in units of 512B-block.  We carefully designed the DMA so that it keeps track of the detailed page swap progress, to be specific, the address range of sub-blocks that have been transferred. When a memory request is targeted at the page being swapped, we use the swap progress indicator to decide where to redirect the memory requests. For instance, if it's a read request and the targeting cache-line falls behind the current swap progress, i.e., the requested data has been transferred to the new location, the DMA redirects the memory request to the destination device.  We spent considerable time to design and verify the logic design to ensure all possible cases are covered and processed properly.
\subsection{PCIe BAR Memory Mapping}
In order to have the applications run on PCIe attached memory, we first need to map the memory space into system address space.
PCIe devices have a set of Base Address Registers (BAR), which are assigned by the platform firmware during the booting process. These BARs carry the base address where the PCIe device is mapped in the system memory-mapped or IO-mapped space. Here we chose the memory-mapped mode as it supports prefetching on our PCIe attached memory. The size of each BAR was programmed into the PCIe IP before FPGA project compilation. \par However, some embedded system might not have enough free system address space for our PCIe memories, which is usually larger than 2GB.
Hence we need to adjust the system mapping appropriately in the device tree of firmware image (e.g. U-boot), to reserve enough address range for the required sizes of PCIe devices. 
\subsection{Arbitrary Latency Cycles}
\label{mimic}
A great advantage of our platform is the flexibility of emulating different memory technologies, by adding arbitrary stall cycles to the access latency. This is essential when you don't have access to a real NVM DIMM, which is the case for many architecture research studies.\par
There are multiple blocks in the workflow (Fig.~\ref{fig:flow}) where you can insert extra cycles of delay, e.g., between the control logic and MC, or on the data return path. Here we use 3D Xpoint as an example to show how we calculated the number of stalling cycles to be inserted.\par 
We measured the round trip time in FPGA cycles to access external DRAM
DIMM first, and then scaled the number of stall cycles according to
the speed ratio between DRAM and 3D Xpoint, as described
in the Table~\ref{tab:nvms}.Thus we have one DRAM DIMM running at
full speed and the other DRAM DIMM emulating the approximate speed of 3D XPoint. Hence the
platform is not constrained to any specific type of NVM, but rather
allows us to study and compare the behaviors across any arbitrary
combinations of hybrid memories. 
\subsection{Driver and Memory Allocator}
\begin{figure}[h]
\centerline{\includegraphics[width=\columnwidth]{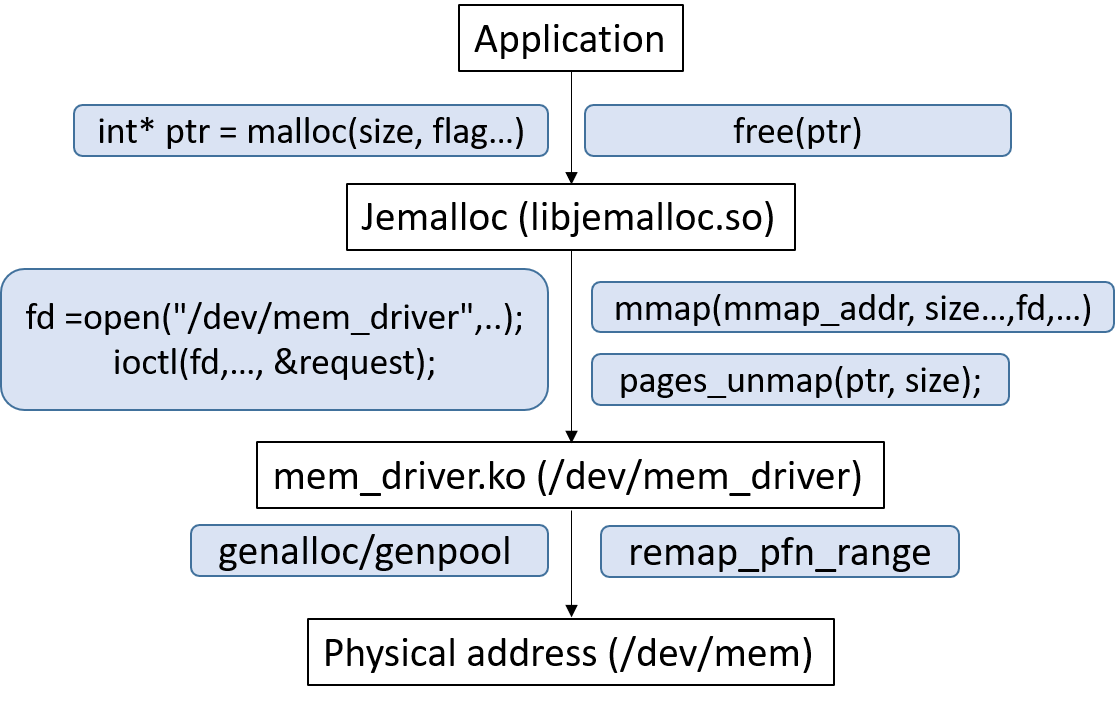}}
\caption{Middleware}
\label{fig:middle}
\end{figure}
The PCIe BAR mapping renders the targeted hybrid memories as a continuous range in the system physical address space, which is presented as a device file \texttt{/dev/mem}.
To mandate the application to only run on the hybrid memories, we need middleware as shown in Fig.~\ref{fig:middle}:
\begin{enumerate}
    \item The driver (\texttt{mem\_driver.ko}) manages the physical frames of the hybrid memories (\texttt{/dev/mem}), with the help of the kernel's \texttt{genpool} subsystem. It sets up the page table using the \texttt{remap\_pfn\_range} kernel function.
   \item We modify the \texttt{pages.c} of \texttt{jemalloc} allocator~\cite{jemalloc}, and use the \texttt{mmap} function to enforce the application allocations within the address range of the specified device file (\texttt{/dev/mem\_driver}).
\end{enumerate}
Users are free to implement their own optimized strategies of virtual memory management inside the allocator, or physical frames management inside the driver. For example, we extended the \texttt{malloc} API, to accept users' hints of memory device preference regarding data placement, and populate these information through the stack to the hardware hybrid memory controller. 
\pgfplotsset{
    /pgfplots/ybar legend/.style={
    /pgfplots/legend image code/.code={%
       \draw[##1,/tikz/.cd,yshift=-0.25em]
        (0cm,0cm) rectangle (12pt, 1em);},
   },
}
\pgfplotstableread{
application	fpga	gem5	champsim
500.perlbench	1.951451546	73839.82973	16418.9097
505.mcf	15.36232386	31079.28459	4890.100918
508.namd	6.353220866	49263.83823	12972.40447
519.lbm	5.685779049	39090.61064	14428.22929
520.omnetpp	1.985561326	10995.30433	6418.788826
523.xalancbmk	4.900496468	13177.75896	6726.882922
525.x264	4.501875102	18237.17124	3877.859354
531.deepsjeng	1.663266195	48400.49791	9251.427302
538.imagick	1.166819817	21169.82902	4064.916096
541.leela	1.275545735	35870.07482	6850.625276
544.nab	1.175347568	37875.96734	8418.632861
557.xz	7.401716196	25630.0926	3809.654436
Geomean	3.167911073	29397.82307	7241.402645
}\slowdown

\pgfplotstableread{
application	rds	wrs
500.perlbench	91269070848	85151563776
505.mcf	3104449445888	3103576014848
508.namd	8357298176	8380891136
519.lbm	271550382080	272163438592
520.omnetpp	257153286144	250019299328
523.xalancbmk	693449916416	696510857216
525.x264	39498645504	39774453760
531.deepsjeng	61425254400	58407272448
538.imagick	4803395584	4827742208
541.leela	102015385600	103786201088
544.nab	80701865984	81461706752
557.xz	44052758528	43823579136
}\requests

\section{Evaluation}
\label{sec:eval}
We implemented an emulation platform that's targeted at the ARM-based mobile computing hybrid memory system. Mobile systems have limited energy budgets, and they're more sensitive to memory power consumption. Hence, we believe the NVM's advantage of minimal standby power will be highly valued.\par 
In this section,
we first present the hardware implementation, and the benchmark software applications. Then
we evaluate the performance of our proposed emulation scheme, by comparing against the two popular software-based simulators, namely Gem5~\cite{gem5} and Champsim~\cite{champsim}. Finally we analyze and discuss some of the
more interesting data points.

\subsection{Methodology}
\label{sec:method}
\subsubsection{Platform Implementation}
\begin{figure}[!hbt]
\centerline{\includegraphics[width=\columnwidth]{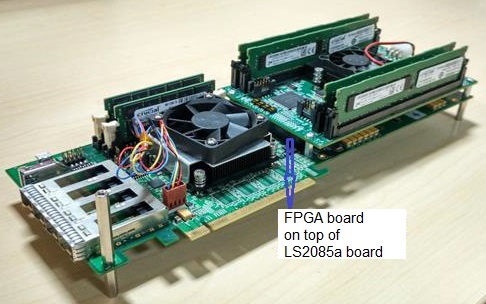}}
\caption{Hardware Boards Setup}
\label{fig:boards}
\end{figure}

\begin{figure}[!hbt]
\centerline{\includegraphics[width=\columnwidth]{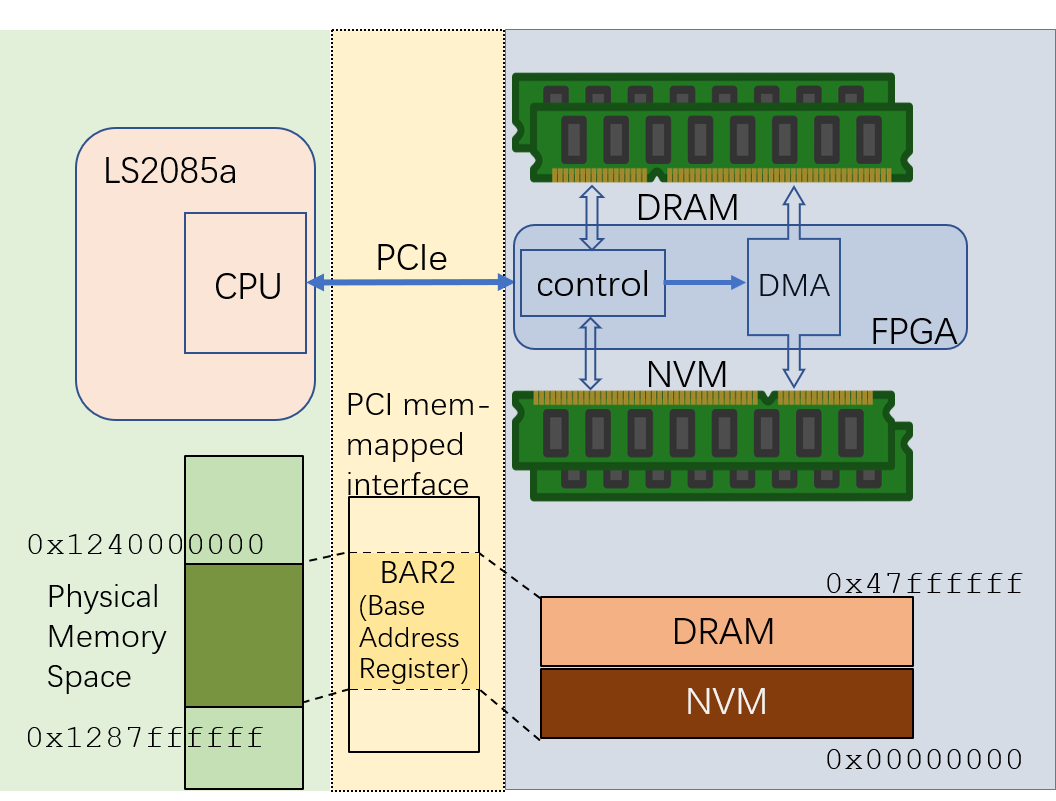}}
\caption{System Implementation}
\label{fig:implementation}
\end{figure}

The system implementation is illustrated in Figure~\ref{fig:implementation}.
We used the customized LS2085A board produced by Freescale which has 8 ARM Cortex A57 cores. The LS2085a connects to the FPGA board via a high-speed PCI Express 3.0 link, and manages the two
memory modules (DRAM and NVM) directly. The platform is physically set up as shown in Figure~\ref{fig:boards}.
Since 3D Xpoint is the only NVM technology that has turned into mass-production (Intel Optane series), we chose it for the presented experiments. We emulated the 3D XPoint device with regular DRAM DIMM, using the approach explained in~\ref{mimic}.\par
The DRAM and NVM memories are
mapped to the address range $[0x1240000000, 0x1288000000)$ of the physical memory space via the PCIe BAR (Base Address
Register) window. Note that the CPU caching is still enabled on the mapped memory space, due to the PCIe BAR configurations.

The
detailed system specification is listed in Table~\ref{tab:System
  Spec}.
\begin{table}[hpbt]
\centering
\settowidth\tymin{\textbf{Component}}
\caption{Emulation System Specification}
\begin{tabulary}{\columnwidth}{L|L}
  \hline
  {\textbf{Component}} & Description \\
  \hline
  CPU & ARM Cortex-A57 @ 2.0GHz, 8 cores, ARM v8 architecture\\
  \hline
  L1 I-Cache & 48 KB instruction cache, 3-way set-associative\\
  \hline
  L1 D-Cache & 32 KB data cache, 2-way set-associative\\
  \hline
  L2 Cache & 1MB, 16-way associative, 64KB cache line size \\
  \hline
  Interconnection & PCI Express Gen3 (8.0 Gbps) \\
  \hline
   DRAM & 128MB DDR4\\
   \hline
   NVM & 1GB 3DXpoint(emulated by DDR4 with added latency) \\
\hline
OS & Linux version 4.1.8 \\
\hline
\end{tabulary}
\label{tab:System Spec}
\end{table}

\subsubsection{Workloads}
We choose applications from the latest version SPEC CPU 2017 benchmark suite~\cite{SPEC_Official}.  To emulate memory pressures of future mobile applications, we
prefer those SPEC CPU 2017 benchmarks which require a larger
working set than the DRAM size in our system.  The details of tested benchmark applications are
listed in Table~\ref{tab:Benchmarks}.

\begin{table}[hbt]
\centering
\settowidth\tymin{\textbf{Memory Footprint}}
\caption{Tested Workloads of SPEC 2017\cite{SPEC_Official}} \label{tab:Benchmarks}

\begin{tabulary}{\columnwidth}{L|L|L}
\hline
{\textbf{Benchmark}} & Description & Memory footprint \\
\hline
\multicolumn {3}{c}{Integer Applications} \\
\hline
500.perlbench	& Perl interpreter & 202MB \\
\hline
505.mcf & Vehicle route scheduling & 602MB\\
\hline
508.namd & Molecular dynamics & 172MB \\
\hline
520.omnetpp	& Discrete Event simulation - computer network	& 241MB \\
\hline
523.xalancbmk	& XML to HTML conversion via XSLT & 481MB \\
\hline
525.x264	& Video compressing & 165MB \\
\hline
531.deepsjeng	& Artificial Intelligence: alpha-beta tree search (Chess) & 700MB \\
\hline
541.leela	& AI: Monte Carlo tree search & 22MB \\
\hline
557.xz & General data compression & 727MB \\
\hline
\multicolumn {3}{c}{Float Point Applications} \\
\hline
519.lbm & Fluid dynamics & 410MB \\
\hline
538.imagick & Image Manipulation & 287MB \\
\hline
544.nab & Molecular Dynamics & 147MB \\
\hline
\end{tabulary}
\end{table}

\subsubsection{Designs Under Test}
\begin{itemize}
    \item Software-based simulators: Gem5, Champsim:\par
    We run the workloads~\ref{tab:Benchmarks} in Gem5 (SE mode) and Champsim, on a x86 workstation with 2 Six-Core Intel Xeon Processor E5-2643 v3 (20M Cache, 3.40 GHz), and 128GB DDR4 Memory. 
    We normalized the simulation time against the running time of native execution on the same machine.
    \item Our proposed FPGA-based emulation platform\par
    We run the same set of workloads on our emulation system~\ref{tab:System Spec}, then we normalized the run time against the time taken by native execution.
    Note that in native execution, the applications run in the on-board DDR4 (16GB) of the LS2085a board.
\end{itemize}

\subsection{Results}
\begin{figure*}[htb]
  \centering
  \includegraphics{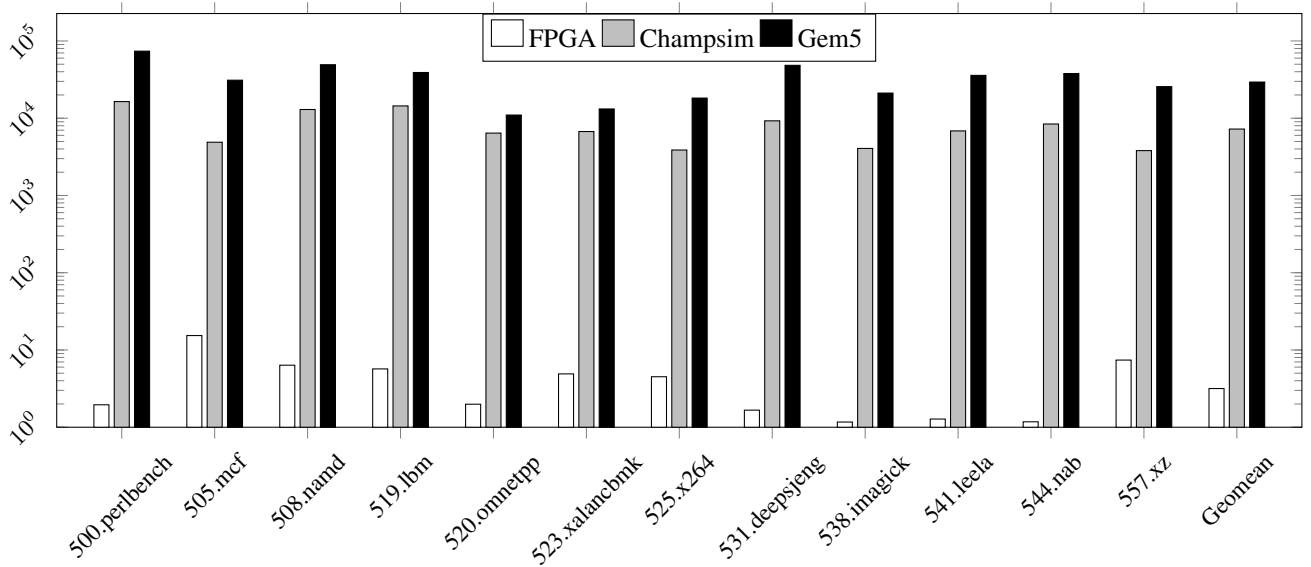}
  \caption{Simulation Time Normalized against Native Execution}
  \label{fig:slowdown}
\end{figure*}

\begin{figure}[htb]
  \centering
  \includegraphics{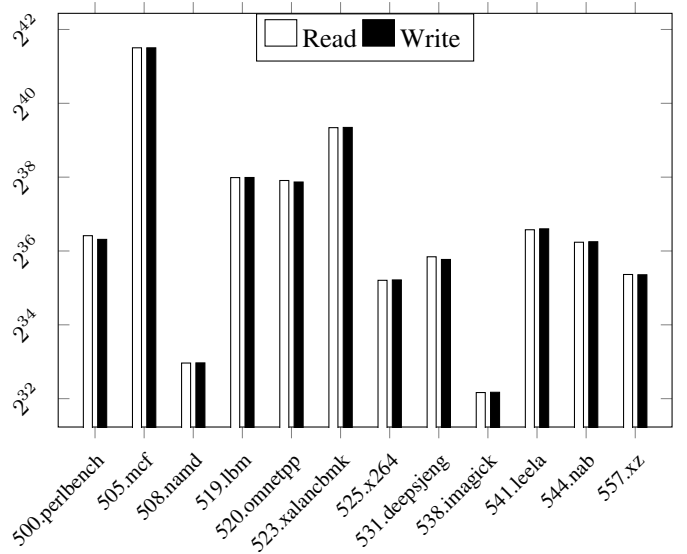}
  \caption{Memory Requests (Bytes)}
  \label{fig:requests}
\end{figure}
Figure~\ref{fig:slowdown} shows the simulation time of each methods, normalized against the run time of native execution.
The geometric mean of slowdown factors indicates the overall time efficiency of each tested design (lower the better).\par
Our FPGA-based emulation system merely experienced 3.17x slowdown versus the native execution, while Gem5 showed 29397.82x slowdown, and Champsim yielded 7241.4x.
Among all the tested workloads, 538.imagick had smallest slowdown (1.17x), and 505.mcf suffered the largest performance penalty (15.36x).\par
To investigate the memory behaviors of these applications, we implemented a variety of performance counters in the FPGA. We used one set of the counters to collect numbers of memory requests generated by each application, as presented in Figure~\ref{fig:requests}. We found that 538.imagick created the fewest memory requests(Read: 4.47GB, Write: 4.49GB); while 505.mcf incurred the most requests (Read: 2.83TB, Write: 2.82TB). Our observation is also confirmed by~\cite{spec2017-workload}, which reports that 505.mcf has the highest cache miss rate, and 538.imagic has the lowest L2/L3 cache miss rates among the tested workloads.
Based on our experience developing the emulation system, we presume the major impact comes from the latency of the PCIe links. Consequently, applications with higher volume of memory accesses will be affected negatively. 
\section{Conclusions}
\label{sec:conc}
Emerging non-volatile memory (NVM) technologies provide higher
capacity and less static power consumption than traditional DRAM. Hybrid memories comprised of both NVM and DRAM
promise to mitigate the energy limits on mobile computing/embedded system. The design space for hyrid memory management pollies are broad, with many
possibilities for HW/SW cross-layer data allocation and migration.
Evaluating hybrid memory systems requires full-stack system simulation involving bus/interconnection, memory controllers, memory devices, etc. With such level of complexity, cycle-level simulation becomes  prohibitive slow for traditional simulation-based techniques.\par
  Here we propose an FPGA based platform for rapid, accurate
hybrid memory system evaluation.  In this platform, only the hybrid
memory management policies themselves are emulated in the FPGA
hardware, the other system components (i.e. the processor, DRAM, etc)
are all real components.  Thus users can focus on the core design and don't have to bother modelling system environments.
Experiments on SPEC 2017 show that our design is 2286 times faster than Champsim, and 9280 times faster than Gem5. \par
The flexibility of FPGA allow users to build performance counter of various functions, providing rich information and new scopes to design exploration.
\bibliographystyle{IEEEtran}
\bibliography{refs.bib}

\begin{thebibliography}{10}
\providecommand{\url}[1]{#1}
\csname url@samestyle\endcsname
\providecommand{\newblock}{\relax}
\providecommand{\bibinfo}[2]{#2}
\providecommand{\BIBentrySTDinterwordspacing}{\spaceskip=0pt\relax}
\providecommand{\BIBentryALTinterwordstretchfactor}{4}
\providecommand{\BIBentryALTinterwordspacing}{\spaceskip=\fontdimen2\font plus
\BIBentryALTinterwordstretchfactor\fontdimen3\font minus
  \fontdimen4\font\relax}
\providecommand{\BIBforeignlanguage}[2]{{%
\expandafter\ifx\csname l@#1\endcsname\relax
\typeout{** WARNING: IEEEtran.bst: No hyphenation pattern has been}%
\typeout{** loaded for the language `#1'. Using the pattern for}%
\typeout{** the default language instead.}%
\else
\language=\csname l@#1\endcsname
\fi
#2}}
\providecommand{\BIBdecl}{\relax}
\BIBdecl

\bibitem{intel-3dxpoint}
{INTEL CORPORATION}, ``Intel optane technology,'' 2016,
  \url{https://www.intel.com/content/www/us/en/architecture-and-technology/intel-optane-technology.html}.

\bibitem{memristor}
K.~Eshraghian, K.-R. Cho, O.~Kavehei, S.-K. Kang, D.~Abbott, and S.-M.~S. Kang,
  ``Memristor mos content addressable memory (mcam): Hybrid architecture for
  future high performance search engines,'' \emph{Very Large Scale Integration
  (VLSI) Systems, IEEE Transactions on}, vol.~19, no.~8, pp. 1407--1417, Aug
  2011.

\bibitem{PCM}
S.~Raoux, G.~W. Burr, M.~J. Breitwisch, C.~T. Rettner, Y.~. Chen, R.~M. Shelby,
  M.~Salinga, D.~Krebs, S.~. Chen, H.~. Lung, and C.~H. Lam, ``Phase-change
  random access memory: A scalable technology,'' \emph{IBM Journal of Research
  and Development}, vol.~52, no. 4.5, pp. 465--479, July 2008.

\bibitem{techinsights}
J.~Choe, ``Intel 3d xpoint memory die removed from intel optane pcm,'' 2017,
  \url{https://bit.ly/3edy4Ev}.

\bibitem{NVM1}
\BIBentryALTinterwordspacing
A.~Chen, ``A review of emerging non-volatile memory (nvm) technologies and
  applications,'' \emph{Solid-State Electronics}, vol. 125, pp. 25--38, 2016.
  [Online]. Available:
  \url{http://www.sciencedirect.com/science/article/pii/S0038110116300867
  https://ac.els-cdn.com/S0038110116300867/1-s2.0-S0038110116300867-main.pdf?_tid=ca6cf681-9266-4eab-b2b1-b5fc588a1cbc&acdnat=1533662521_dc98aa6ba5b7f6da59bc748a06b160b0}
\BIBentrySTDinterwordspacing

\bibitem{NVM2}
\BIBentryALTinterwordspacing
S.~Mittal and J.~S. Vetter, ``A survey of software techniques for using
  non-volatile memories for storage and main memory systems,'' \emph{IEEE
  Transactions on Parallel and Distributed Systems}, vol.~27, no.~5, pp.
  1537--1550, 2016. [Online]. Available:
  \url{https://ieeexplore.ieee.org/ielx7/71/7449033/07120149.pdf?tp=&arnumber=7120149&isnumber=7449033}
\BIBentrySTDinterwordspacing

\bibitem{yang:2012}
J.~J. Yang, D.~B. Strukov, and D.~R. Stewart, ``Memristive devices for
  computing,'' \emph{Nature Nanotechnology}, Dec 2012.

\bibitem{nvm_price}
A.~Shilov, ``Pricing of intel's optane dc persistent memory modules,'' 2019,
  \url{https://www.anandtech.com/show/14180/pricing-of-intels-optane-dc-persistent-memory-modules-leaks}.

\bibitem{champsim}
ChampSim, ``Champsim,'' 2016, \url{https://github.com/ChampSim/ChampSim}.

\bibitem{gem5}
\BIBentryALTinterwordspacing
N.~Binkert, B.~Beckmann, G.~Black, S.~K. Reinhardt, A.~Saidi, A.~Basu,
  J.~Hestness, D.~R. Hower, T.~Krishna, S.~Sardashti, R.~Sen, K.~Sewell,
  M.~Shoaib, N.~Vaish, M.~D. Hill, and D.~A. Wood, ``The gem5 simulator,''
  \emph{SIGARCH Comput. Archit. News}, vol.~39, no.~2, pp. 1--7, Aug. 2011.
  [Online]. Available: \url{http://doi.acm.org/10.1145/2024716.2024718}
\BIBentrySTDinterwordspacing

\bibitem{7155440}
T.~Nowatzki, J.~Menon, C.~Ho, and K.~Sankaralingam, ``Architectural simulators
  considered harmful,'' \emph{IEEE Micro}, vol.~35, no.~6, pp. 4--12, Nov 2015.

\bibitem{Hassan:2015}
\BIBentryALTinterwordspacing
A.~Hassan, H.~Vandierendonck, and D.~S. Nikolopoulos, ``Software-managed
  energy-efficient hybrid dram/nvm main memory,'' in \emph{Proceedings of the
  12th ACM International Conference on Computing Frontiers}, ser. CF '15.\hskip
  1em plus 0.5em minus 0.4em\relax New York, NY, USA: ACM, 2015, pp.
  23:1--23:8. [Online]. Available:
  \url{http://doi.acm.org/10.1145/2742854.2742886}
\BIBentrySTDinterwordspacing

\bibitem{span}
\BIBentryALTinterwordspacing
V.~Fedorov, J.~Kim, M.~Qin, P.~V. Gratz, and A.~L.~N. Reddy, ``Speculative
  paging for future nvm storage,'' in \emph{Proceedings of the International
  Symposium on Memory Systems}, ser. MEMSYS '17.\hskip 1em plus 0.5em minus
  0.4em\relax New York, NY, USA: ACM, 2017, pp. 399--410. [Online]. Available:
  \url{http://doi.acm.org/10.1145/3132402.3132409}
\BIBentrySTDinterwordspacing

\bibitem{Liu:2017}
\BIBentryALTinterwordspacing
H.~Liu, Y.~Chen, X.~Liao, H.~Jin, B.~He, L.~Zheng, and R.~Guo,
  ``Hardware/software cooperative caching for hybrid dram/nvm memory
  architectures,'' in \emph{Proceedings of the International Conference on
  Supercomputing}, ser. ICS '17.\hskip 1em plus 0.5em minus 0.4em\relax New
  York, NY, USA: ACM, 2017, pp. 26:1--26:10. [Online]. Available:
  \url{http://doi.acm.org/10.1145/3079079.3079089}
\BIBentrySTDinterwordspacing

\bibitem{ramos}
L.~E. Ramos, E.~Gorbatov, and R.~Bianchini, ``Page placement in hybrid memory
  systems,'' in \emph{Proceedings of the International Conference on
  Supercomputing}, ser. ICS ’11.\hskip 1em plus 0.5em minus 0.4em\relax New
  York, NY, USA: Association for Computing Machinery, 2011, p. 85–95.

\bibitem{CSu}
\BIBentryALTinterwordspacing
C.~Su, D.~Roberts, E.~A. Le\'{o}n, K.~W. Cameron, B.~R. de~Supinski, G.~H. Loh,
  and D.~S. Nikolopoulos, ``Hpmc: An energy-aware management system of
  multi-level memory architectures,'' in \emph{Proceedings of the 2015
  International Symposium on Memory Systems}, ser. MEMSYS '15.\hskip 1em plus
  0.5em minus 0.4em\relax New York, NY, USA: Association for Computing
  Machinery, 2015, p. 167–178. [Online]. Available:
  \url{https://doi.org/10.1145/2818950.2818974}
\BIBentrySTDinterwordspacing

\bibitem{Chung:PROTOFLEX}
\BIBentryALTinterwordspacing
E.~S. Chung, E.~Nurvitadhi, J.~C. Hoe, B.~Falsafi, and K.~Mai, ``A
  complexity-effective architecture for accelerating full-system multiprocessor
  simulations using fpgas,'' in \emph{Proceedings of the 16th International
  ACM/SIGDA Symposium on Field Programmable Gate Arrays}, ser. FPGA '08.\hskip
  1em plus 0.5em minus 0.4em\relax New York, NY, USA: Association for Computing
  Machinery, 2008, p. 77–86. [Online]. Available:
  \url{https://doi.org/10.1145/1344671.1344684}
\BIBentrySTDinterwordspacing

\bibitem{Fytraki:ReSim}
S.~{Fytraki} and D.~{Pnevmatikatos}, ``Resim, a trace-driven, reconfigurable
  ilp processor simulator,'' in \emph{2009 Design, Automation Test in Europe
  Conference Exhibition}, 2009, pp. 536--541.

\bibitem{Aport}
M.~Pellauer, M.~Vijayaraghavan, M.~Adler, Arvind, and J.~Emer, ``A-ports: An
  efficient abstraction for cycle-accurate performance models on fpgas,'' in
  \emph{Proceedings of the 16th International ACM/SIGDA Symposium on Field
  Programmable Gate Arrays}, ser. FPGA '08.\hskip 1em plus 0.5em minus
  0.4em\relax New York, NY, USA: Association for Computing Machinery, 2008, p.
  87–96.

\bibitem{Sunwoo:PrEsto}
D.~{Sunwoo}, G.~Y. {Wu}, N.~A. {Patil}, and D.~{Chiou}, ``Presto: An
  fpga-accelerated power estimation methodology for complex systems,'' in
  \emph{2010 International Conference on Field Programmable Logic and
  Applications}, 2010, pp. 310--317.

\bibitem{Njoroge:ATLAS}
N.~{Njoroge}, J.~{Casper}, S.~{Wee}, Y.~{Teslyar}, D.~{Ge}, C.~{Kozyrakis}, and
  K.~{Olukotun}, ``Atlas: A chip-multiprocessor with transactional memory
  support,'' in \emph{2007 Design, Automation Test in Europe Conference
  Exhibition}, 2007, pp. 1--6.

\bibitem{jemalloc}
J.~Evans, ``Jemalloc,'' 2016, \url{http://jemalloc.net/}.

\bibitem{SPEC_Official}
SPEC, ``{SPEC CPU2017 Documentation},'' 2017,
  \url{https://www.spec.org/cpu2017/Docs/}.

\bibitem{spec2017-workload}
\BIBentryALTinterwordspacing
A.~Limaye and T.~Adegbija, ``A workload characterization of the spec cpu2017
  benchmark suite,'' in \emph{2018 IEEE International Symposium on Performance
  Analysis of Systems and Software (ISPASS)}, 2018, Conference Proceedings, pp.
  149--158. [Online]. Available:
  \url{https://ieeexplore.ieee.org/ielx7/8360838/8366921/08366949.pdf?tp=&arnumber=8366949&isnumber=8366921&ref=}
\BIBentrySTDinterwordspacing

\end{thebibliography}
\end{document}